\documentclass[acmlarge]{acmart}

\usepackage{booktabs} 
\usepackage{ulem}

\usepackage[ruled]{algorithm2e} 
\usepackage{bytefield}

\acmJournal{IMWUT}
\acmVolume{9}
\acmNumber{4}
\acmArticle{39}
\acmYear{2017}
\acmMonth{11}

\setcopyright{acmcopyright}

\acmDOI{0000001.0000001}

\begin{document}
\title{CryptoCam: Privacy Conscious Open Circuit Television}

\author{Gerard Wilkinson}
\affiliation{
  \institution{Open Lab, Newcastle University}
  \city{Newcastle upon Tyne}
  \country{UK}
}
\author{Dan Jackson}
\affiliation{
  \institution{Open Lab, Newcastle University}
  \city{Newcastle upon Tyne}
  \country{UK}
}
\author{Andrew Garbett}
\affiliation{
  \institution{Open Lab, Newcastle University}
  \city{Newcastle upon Tyne}
  \country{UK}
}
\author{Reuben Kirkham}
\affiliation{
  \institution{Open Lab, Newcastle University}
  \city{Newcastle upon Tyne}
  \country{UK}
}
\author{Kyle Montague}
\affiliation{
  \institution{Open Lab, Newcastle University}
  \city{Newcastle upon Tyne}
  \country{UK}
}
\email{g.wilkinson@newcastle.ac.uk}
\email{dan.jackson@newcastle.ac.uk}
\email{andy.garbett@newcastle.ac.uk}
\email{reuben.kirkham@newcastle.ac.uk}
\email{kyle.montague@newcastle.ac.uk}

\renewcommand\shortauthors{Wilkinson, G. et al.}

\begin{abstract}
The prevalence of Closed Circuit Television (CCTV) in today's society has given rise to an inherent asymmetry of control between the watchers and the watched. A sense of unease relating to the unobservable observer (operator) often leads to a lack of trust in the camera and its purpose, despite security cameras generally being present as a protective device. In this paper, we detail our concept of Open Circuit Television and prototype CryptoCam, a novel system for secure sharing of video footage to individuals and potential subjects nearby. Utilizing point-of-capture encryption and wireless transfer of time-based access keys for footage, we have developed a system to encourage a more open approach to information sharing and consumption. Detailing concerns highlighted in existing literature we formalize our over-arching concept into a framework called Open Circuit Television (OCTV). Through CryptoCam we hope to address this asymmetry of control by providing subjects with data equity, discoverability and oversight.
\end{abstract}

%
%


%
%


\maketitle

\section{Introduction}
Video cameras are versatile tools, used for a plethora of purposes, from capturing one's day to day life (lifelogging), as an assistive technology \cite{Hodges2006SenseCam:Aid,Marcu2012Parent-drivenSupport}, a means for sousveillance \cite{Mann2002Sousveillance:Environments.}, as a tool for documenting human rights abuses \cite{Gregory2010CamerasConsent}, as a scientific tool \cite{Zhang2007Video-BasedCameras}, as a form of sensing (e.g. Kinect\footnote{Kinect, https://developer.microsoft.com/en-us/windows/kinect}), for recording current affairs (for journalism), or as a tool for enhancing security in public spaces. Yet, many existing surveillance systems are closed in nature: control often resides with the organization which has recorded the footage, or to put it another way, with the watcher rather than the watched. For the ordinary citizen, the process for obtaining footage can be challenging. One would have to notice the camera (many are obscured), determine who owns it and in turn, make a formal request through legalistic processes. The result might be that someone is provided with the footage several months later on a DVD (in a format and timing that might well be of little use).

The closed nature of surveillance cameras, along with other privacy challenges, such as that video footage can sometimes contain (directly or indirectly) sensitive personal data, and the desire for individual privacy more generally, leads to concern over their configuration. There is a pressing need for cameras, but also a similar need to configure these systems in a way that does not overly intrude into the rights of those being recorded. The Ubicomp community has adopted a variety of approaches aimed at achieving this, including careful positioning of cameras to ensure that their field of view is only focused on the target of interest \cite{Thomaz2013TechnologicalCameras}. However, discoverability is an under explored area for CCTV, in terms of presence, purpose and configuration.

CryptoCam looks to enable new possibilities for configuration by deploying industry standard encryption technologies to encrypt footage at point of capture with keys to the footage distributed to relevant parties. The emphasis of CryptoCam is in redressing the asymmetry of control and access over footage from cameras. Through moving the locus of control from operators to potential subjects, either in its entirety with a complete local encryption solution with keys only held by subjects, or a more balanced solution with master keys held by the camera operator. A client application on subjects' mobile phone records keys from cameras nearby using Bluetooth, placement and other configuration information can be shared at this time. The process is entirely anonymous with no identification of the subjects needed or warranted. This particular configuration has several advantages: in many settings (within the EU), it would fall within the domestic use exemption of the General Data Protection Regulation (GDPR)\footnote{General Data Protection Regulation, https://eur-lex.europa.eu/legal-content/EN/TXT/?uri=celex:32016R0679} it is likely to enhance user trust (especially in the context of negative perceptions of cameras). We hope its use will be accepted in a broader range of scenarios, while also enhancing consent and accessibility in relation to the footage.

This paper presents a novel concept, CryptoCam, which will be made available as an open source tool that others can build upon and deploy in a wide range of settings. We develop the context in which CryptoCam sits in Section \ref{motivationbackground}, including the development and exploration of how this context intersects with a broad range of use cases. We then move to describe our framework we call Open Circuit Television (OCTV) (Section \ref{octv}), before describing the implementation of CryptoCam itself (Section \ref{cryptocam}) and the particular advantages of how CryptoCam has been configured (Section \ref{discussion}). This paper concludes (Section \ref{conclusion}) with a discussion of CryptoCam, and how it fits into a wider family of potential systems that distribute the control of the data that they record.

\section{Motivation and Background}\label{motivationbackground}
\subsection{Existing CCTV usage}\label{existingcameras}
In recent years we have seen an increase in both the number and application areas of CCTV Cameras, including product security surveillance, workplace safety, dash- and taxi-cams and police body cameras. There has been a wealth of evidence to validate the installation of much of this infrastructure \cite{Welsh2002CrimeReview,Welsh2003EffectsCrime}. However, the proliferation of CCTV in today's society leads to renewed privacy concerns.

Closed Circuit Television is, by definition, closed to the public. CCTV subjects themselves are typically the ones excluded, leading to apprehensions about being covertly, and perhaps maliciously surveiled. This self-defeating prejudice about CCTV often prevents adoption of cameras for their original intended purpose as a deterrent and protective device. The \textit{`Big-Brother'} attitude of governments and organizations in utilizing CCTV for other means, including enforcement and performance monitoring, has again reduced CCTV to more of a nuisance to subjects rather than an asset. Subjects can be frustrated further by the often complex and arduous processes to retrieve footage of themselves. Research into breaking down some of these barriers to access and discoverability can open up potential application scenarios for CCTV and other video recording technologies. Moncrieff and Mollers argue that dynamism in privacy controls is the answer to preserving privacy of subjects while maintaining the purpose of the cameras and that designers should build systems which integrate privacy controls \cite{Moncrieff2009DynamicSurveillance,Mollers2013PrivacyGermany}. \textit{``Privacy as an optimization problem''} storing the minimum amount of information about subjects as possible, \textit{``data hiding''} with sensitive parts of footage obscured or removed, \textit{``context-aware surveillance''} with a dynamic privacy policy based on the current state of the camera or subjects and finally \textit{``data equity''} with footage more open to relevant individuals.

Existing work on dynamism in camera technology has focused on \textit{``data hiding''} and \textit{``optimizing''} using Computer Vision techniques to hide irrelevant or sensitive parts of an image \cite{Senior2005EnablingVision,senior2003blinkering,Chattopadhyay2007PrivacyCam:DSP,Boult2005PICO:Obscuration}. These approaches to enhancing privacy provide a flexible and convenient means to protect subjects, however, the vision based techniques are limited by what they can accurately detect and \textit{`understand'}. Privacy Cam \cite{Chattopadhyay2007PrivacyCam:DSP}, could detect individuals and encrypt their portion of the image with a unique key for later decryption by the subject. However, beyond distinguishing between people, user identification of an individual person was not performed, nor was distribution of users' key.

Goold \textit{et al.} discusses the \textit{``unobservable observer''} \cite{Goold2002PrivacyObserver}, emphasizing one of the key concerns of individuals concerning CCTV \cite{Smith1996InformationPractices,Nguyen2011SituatingRecording}. The remote observer of footage is largely unknown. It is unclear the status of any particular camera: for example, who has access to the camera feed? Is it continuously monitored? Recorded? How long is it stored for? Abuse in CCTV operations is also a cause for concern with regards to privacy \cite{Taylor2014CCTV:Effectiveness}. The asymmetry of power and control creates a potential for abuse. Increasingly, private firms are \textit{``donating''} their footage to state services, with camera operator, commercial organized and trained \cite{Walby2005Open-StreetCanada}. Oversight in the small application of CCTV is problematic. The increasing usage of CCTV is in the trust of security cameras as such cameras are deployed in more sensitive areas (e.g. schools \cite{Taylor2012TheSchool, Yorke2010TeachersSchools}). How can regulation and technical solutions keep pace to protect privacy?

Freedom of Information requests and other relevant legislation governing sharing of footage and signage for CCTV have failed to keep pace with advancing technology. Current processes are evidenced as being inadequate \cite{Spiller2015GainingBe}, obtaining existing footage can be challenging. One would have to notice the camera (many are obscured), then determine who owns it before, in turn, making a formal request through legalistic processes. The result might be that someone is provided with the footage several months later on a DVD. This process if often inadequate and inefficient. Technical approaches to support discovery in terms of presence, access to data and configuration would potentially lead to a greater level of trust in CCTV and better practices for the use of CCTV.

\subsection{Configuration for Privacy}
Privacy is an inherently subjective term, indeed Westin states \textit{``no definition of privacy is possible''} \cite{Westin1966SciencePrivacy}. Governments and corporations often offer assurances of their attitude towards preserving this fundamental concept in modern society. Arguably, any such privacy measure in a system is only as secure as the trustworthiness of the engineers involved in its construction. Industry has, in response, attempted to move control and responsibility for the protection of our personal data to the user. This reduces (but does not eliminate) the trust requirement for engineers. WhatsApp\footnote{WhatsApp, https://www.whatsapp.com}, iMessage\footnote{Apple iMessage, https://support.apple.com/en-gb/explore/messages}, iCloud\footnote{Apple iCloud, https://www.icloud.com} and Keybase\footnote{Keybase, https://keybase.io} have all migrated their encryption key handling to the end user. The user login is used to decrypt the keys for each service. Mozilla Firefox have a comprehensive example of this with their accounts data\footnote{Firefox Accounts, https://www.mozilla.org/en-US/privacy/firefox/}. However, this is compromised by government interventions in existing cryptographically secure implementation of services, requiring the inclusion of back doors into systems making them fundamentally insecure \cite{Griffin2015WhatsAppPlans}.

Westin reasons that privacy is \textit{``the claim of individuals, groups, or institutions to determine for themselves when, how, and to what extent information about them is communicated to others''}. Therefore, placing the keys to our data in the hands of the users, is a move towards the desirable goal of increased privacy. However, oversight in such cases is still limited, with the type of data held and its usage still largely obscured. Companies like Google have demonstrated a cavalier attitude to privacy, with rather objectionable abuse of their advertising and search powers resulting in large fines from governments \cite{Boffey2017GoogleResults} and consumer law suits \cite{Fadilpasic2017UKAbuse}. Provision of services for free can often be accompanied by aggressive collection of user data to fund these services. Moving control of the data to the users, provides and interesting paradigm shift in the approach to management and control over data and its storage. BitTorrent protocol\footnote{BitTorrent, https://en.wikipedia.org/wiki/BitTorrent} and IPFS\footnote{IPFS, https://ipfs.io} are the most widely used implementations to date of peer supported distribution and storage networks respectively. Indeed, Yeung \textit{et al.} argue for such an approach: for social networks with user data stored in a decentralized manner \cite{Yeung2009DecentralizationNetworking}. The wealth of video and data recorded of individuals is vast and ever growing. Adequate access, oversight and control procedures have failed to keep pace.

Commonly used technical means of limiting capture include reducing the resolution of the camera \cite{Jackson2009FiberBoard}, changing the framerate (including to on the order of minutes \cite{Hodges2006SenseCam:Aid}), or changing the frequency spectrum (e.g. room occupancy\footnote{Density, https://www.density.io}). More technically advanced approaches involve processing the content of the video, for instance by blurring facial features, or even by changing what is recorded based upon who is in the picture \cite{Chattopadhyay2007PrivacyCam:DSP,Zhang2016Privacy-friendlySystem}. There are also pragmatic approaches, such as raising awareness through signage, and making the camera itself obvious.

More generally in image capture, careful configuration has been used to allow cameras to be used as a sensor while maintaining privacy. Approaches include positioning, which is often confined with a constrained field of view in order to ensure that the system is only focused on the location or object of interest \cite{Thomaz2013TechnologicalCameras}. Channeling light using optical fibres in tabletop surfaces to perform input \cite{Jackson2009FiberBoard} or fingerprint recognition for user authentication \cite{Holz2013Fiberio}, while not being able to image things above the surface. Similarly, Iris \cite{Montague2017PrototypingSurfaces} achieved this while performing more general object recognition by using an optical diffuser.

\subsection{Summary}
Our work is focused on deploying technology to provoke change in the handling of data collected about the public. We focus our work on video cameras, particularly CCTV. In our work we hope to introduce the concept of Open Circuit Television, with a new approach to discovery, ownership and control over cameras. In doing so we hope to address the asymmetry of control, access and privacy of subjects of CCTV, a currently under-explored area of Ubicomp.

\section{Building a Framework for Open Circuit Television}\label{octv}
In order to understand the possibilities for a more open approach to camera infrastructure we constructed a framework for Open Circuit Television. We developed this framework to guide our research and design with a view to constructing a system that covered the wealth of issues relating to CCTV cameras and the consequences of change in certain aspects of surveillance infrastructure introduced by OCTV.

We consider OCTV as an opportunity to redefine existing constraints of camera configuration through applying novel technological solutions. We highlight five key areas in our framework: space \& configuration, discoverability, access and restrictions, and symmetry in control. This framework aims to guide development of OCTV technological solutions to address the issues we discuss in this work. We also hope to use this framework to highlight areas which users may need to be aware of when considering how to respond to the presence of a camera. The overall focus of this framework is based on the equitable sharing of footage and a more open approach to cameras \cite{Brush2013DigitalWatch}.

\subsection{Configuration \& Space}
The configuration and space in which a camera operates can influence many features of a camera and its corresponding placement. Considering these new parameters for configuration allows us to build upon these in the subsequent components of OCTV. Oversight of these parameters could help to balance the asymmetry in the accessing of data and placement of cameras, and hopefully reduce abusive practices through making this configuration more public. Issues arise around the appropriateness of cameras in certain settings and how such cameras can be adapted to better suit the space and its occupants.

\subsubsection{Public / Private}
Public and private camera settings have different priorities in their considerations. Public spaces are often larger areas and can be more liberal with their data access policies. Subjects within a public space would perhaps be more open as their conduct is already expected to be public. Accidentally granting access to those outside of the view of a camera, yet still near, is perhaps less problematic. Conversely private spaces require more careful consideration of sharing and storage policies as the potential for a breach of privacy is increased.

\subsubsection{Fixed / Mobile}
The positioning and placement of a camera also impacts upon the usage and suitability of deployment. Fixed and mobile cameras have different design considerations. Fixed cameras, for instance, have a more predictable content, the overall area in view is fixed, and is often accompanied with appropriate signage and expectations. Fixed cameras are therefore more appropriate in surveillance and security contexts. Conversely, mobile cameras can provide much more raw and unfiltered footage and the scene is unpredictable. A body-cam, or other mobile camera, may capture unintended and personal information. Subjects can typically make reasonable adjustments before being filmed by a fixed camera, but this is often not so easily the case with mobile cameras. Typically, suitable signage is impractical and the storage and sharing of footage is more difficult to control.

\subsubsection{Lifecycle and Storage}
Visibility of the lifecycle and storage of footage is paramount to trust in a system. How long a camera recording is retained, even if that period is indefinite (no longer permitted in the EU under GDPR), is information that users could be made aware of. The process for destruction, and the circumstances in which footage may be used, and methods by which privacy will be preserved, should all be considered here. Furthermore, the manner in which footage is stored is important: whether encryption technologies are used, and any key storage and sharing solutions. Not only would potential subjects be able to understand how their personal information is stored, but again, to provide external stimulus to best practices.

\subsection{Discoverability}
The ability of users to discover the presence and configuration of cameras in a space is a vital aspect of the oversight and trust-building we look to encourage with Open Circuit Television. Simple, reliable and accurate means of discovery of a camera would be the objective, as it is important that subjects can reflect upon a situation being recorded, and this encourages best practices for operators.

\subsubsection{Responsibility}
Establishing who is responsible for a particular camera is a prevalent issue within existing camera installations. Private and public cameras' existing access procedures are often unclear and can lack signage or clarity around camera and data owner responsibility (despite the requirement of providing this information publicly in many jurisdictions). Clear advertising of this information can potentially speed up such access and reduce administrative work involved in these procedures.

\subsubsection{Location}
The exact location of a camera within a space is again an important consideration in discoverability. Knowing that there is a camera nearby, but not exactly where it is placed, can prevent subjects' preparations towards privacy. Conversely, providing detailed placement information without breaching privacy or security of individuals or buildings is a point of concern. A frame from the camera intended to establish position could unduly leak personal information.

\subsection{Access and Restrictions}
Access to footage, with varying degrees of access offered, forms a key part of OCTV, restricting: the video feed, audio channel and file access. One approach would be to base access level on a subject's appropriateness: those closer to the camera receiving higher-level access, while fringe subjects receive limited access. Footage availability may be delayed for review, made available immediately or withheld by the operator. Restrictions on the usage of, and retrieval of, footage (both for the operator and subject), soliciting permission/release for operators to use footage, subjects re-publishing restrictions, and removal of identifiable information from footage, can all  form part of the camera configuration.

\subsubsection{Recording State}
Visibility over a camera's specific state is an important consideration for OCTV: whether the camera is recording, and whether it is being actively monitored or used for post-reflective surveillance. These factors may depend on a schedule (i.e. only recording for certain times of the day, or only actively monitored at certain times). While such information for many camera situations may be obvious, or otherwise not necessary to disclose, there is potential for disclosing the details and type of surveillance being conducted. In particular, active surveillance is something that again perhaps propagates negative perceptions about cameras, and better visibility of this information could address this.

\subsection{Symmetry in Control}
The current asymmetry of control over footage is a balance that can be addressed in OCTV. Increasing oversight of camera systems, and providing subjects with some aspects of control through novel mechanisms, moves towards breaking down existing control paradigms and more open and democratic cameras.

\section{Scenarios}
To illustrate our motivations for Open Circuit Television we will explore each of these exemplar scenarios. We uncover and identify issues surrounding the ownership and access rights of media, and the discoverability of cameras and footage.

\subsection{Workplace Cameras}\label{workplacecam}
Workplace cameras are commonplace, yet under existing legislation in many jurisdictions, private organizations are under no obligation to share camera footage with their employees \cite{Massimi2010UnderstandingLife,Friedman2006ThePlace}. Instead, they are typically configured primarily for the needs of the owner. As such these cameras are often perceived as being used for organizational surveillance and performance monitoring, making them undesirable to the workforce they could also be protecting.
 
\subsection{Dash- and Taxi Cameras}\label{taxicam}
Taxi cameras are present in many private taxi cabs today, despite the public being less accepting of their presence \cite{YouGov2010SurveillanceStats}, their existence has improved driver safety \cite{Menendez2013EffectivenessRates.}. These protective devices are an inexpensive way for drivers and, to some extent, passengers to record generally irrefutable evidence of what occurred on a journey. However, the footage is controlled and owned by the driver or operator. In many jurisdictions these individuals or organizations are not compelled by law to hand over footage in criminal proceedings. As such, whilst arguably the camera is there for the protection of the driver and passengers; in reality it is the driver that maintains control. In situations where the driver is at fault this ownership model could prove unfair. Furthermore, the footage captured within a taxi can be sensitive in nature due to an assumed privacy from the enclosed space.

\subsection{Body Cameras}\label{bodycams}
Police body cameras are a relatively recent means for police forces to provide adequate oversight over the actions of their officers and provide evidence in criminal proceedings. The presence of these cameras is undoubtedly of benefit to both officers and the general public \cite{Tullio2010ExperienceEngagement}, camera footage is a powerful tool in many court cases. Footage has been used as irrefutable evidence in trials \cite{USSupremeCourtScottCenter}, though its irrefutability is obviously questioned \cite{WassermanVideoHarris}, it is however a very powerful form of evidence \cite{McLAGAN2006Introduction:Public,Gregory2010CamerasConsent}. Recent cases in the US have further demonstrated the importance of body cameras for oversight in police shooting cases, both to exonerate and incriminate officers. However, access to these recordings is restricted, their release determined by police departments, sometimes understandably. In cases of police brutality, there is often little justification for withholding recordings, yet it still happens. While there are already procedures in place in most jurisdictions for subjects to access footage, these processes are often expensive (either to the individual or the tax payer), time consuming and inaccessible to many due to the bureaucracy related to these submissions.

\subsection{Domestic / Home Security Cameras}
Thus far, we have focused on institutional and commercial surveillance systems, however, there are a large number of small private business and domestic cameras installed around the world. New smart cameras have accelerated the numbers of domestic cameras, indeed such cameras are described as effective deterrents against burglars \cite{Smithers2017FormerDeterrent}. Companies such as Nest\footnote{Nest, https://nest.com/} provide features like person detection and event triggering through services like IFTTT\footnote{IFTTT, https://ifttt.com}. There is no register of owners in this category, so it is difficult to accurately assess numbers here, however, there has been a significant increase in these cameras usage in criminal cases \cite{Hill2018HenryDeath}. The discoverability of these cameras is fast becoming a societal problem: with cameras being directed at neighbors' properties \cite{Dominiczak2013CCTV:Neighbours} and tenants/visitors having little awareness of cameras or control over their own privacy.

\subsection{Live Streaming Services}
The raw, unfiltered nature of ``live streaming'' is one of the immediate attractions of these recent platforms. Post-production editing has usually provided the opportunity to make such videos sensitive to other subjects' privacy. Soliciting of consent to record is not legally required in many jurisdictions in public spaces. Soliciting consent during a live stream is impractical in most cases and detracts from the medium's raw nature. Technology to detect and obscure sensitive parts of the footage may prove too aggressive, yet soliciting consent from those nearby via technological means could improve privacy in live streaming.

\section{CryptoCam} \label{cryptocam}
CryptoCam is a prototype Open Circuit Television system, using state of the art encryption technologies and Bluetooth to secure footage and share access tokens with subjects. Combining existing technologies to produce a simple yet powerful system to realize the possibilities for Open Circuit Television.

\begin{figure}
  \includegraphics[height=180pt]{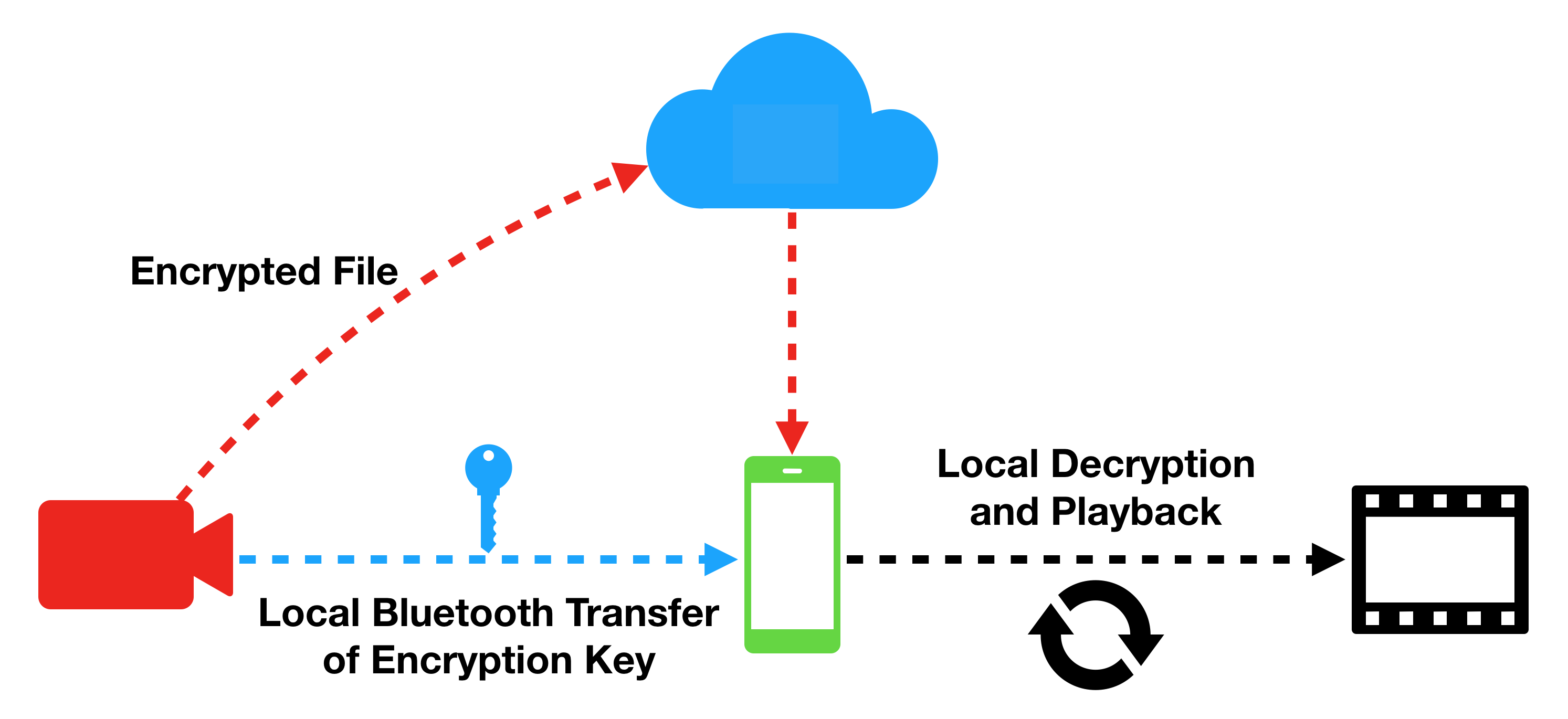}
  \caption{Footage is recorded, encrypted and uploaded to a cloud storage provider. Encryption keys are distributed to phone-based listening clients nearby. Phone clients later retrieve encrypted footage from the cloud and decrypt the contents using the key previously provided locally, then the footage can be played on the device.}
  \label{fig:dataflow}
\end{figure}

CryptoCam is a camera that combines encryption and Bluetooth to create a secure and anonymous footage sharing system. Video footage is taken from the camera and encrypted, the key used for encryption is randomly selected at the beginning of the recording interval. This key is made available to users nearby over Bluetooth (Fig. \ref{fig:packet}). Software clients installed on users' devices listen for encryption keys broadcast from cameras nearby, storing these keys for possible later use. The flow of data is described in Fig. \ref{fig:dataflow}. Cameras also describe a file access protocol, these files can be later retrieved in an encrypted form from the specified location. If required, the encryption keys collected locally are then used to decrypt the encrypted video footage once retrieved. Crucially, all decryption occurs on the device, the key never leaves the device so the user is not identified to the server and the encryption key stays private to the camera and nearby devices. Periodically, the current, encrypted, recording is uploaded to the specified file store and erased locally, and a new encrypted recording is started using another, randomly selected, key.

\subsection{Camera Implementation}
The prototype implementation uses a Raspberry Pi Zero W\footnote{Raspberry Pi Zero W, https://www.raspberrypi.org/products/raspberry-pi-zero-w/} with a Pi Camera v2.1 module\footnote{Raspberry Pi Camera module V2, https://www.raspberrypi.org/products/camera-module-v2/}. This hardware provides an inexpensive, small and low power package which is highly configurable. The CryptoCam software runs on Node.js\footnote{Node.js, https://nodejs.org} and an industry-standard encryption stack (OpenSSL\footnote{OpenSSL, https://www.openssl.org}) is used to handle encryption of footage and generation of keys. The camera is configured to continuously record with a pre-configured segment interval. A random 256-bit key is generated and broadcast to any devices listening nearby, along with a video identifier. At the segment interval boundary the recorded file is hashed (SHA-256), and part of the hash broadcast alongside the next key exchange. This guards against subsequent changes to the recording file (however manipulation could occur before the recording is complete).

\begin{figure}
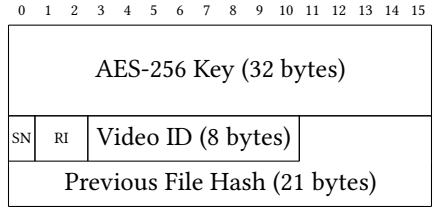

  \begin{bytefield}[bitwidth=1em]{16}
    \bitheader{0-15} \\
    \wordbox{2}{AES-256 Key (32 bytes)} \\
    \bitbox{1}{\tiny SN} & \bitbox{2}{\tiny RI} & 
    \bitbox{8}{Video ID (8 bytes)} & \bitbox[tlr]{5}{} \\
    \wordbox[lrb]{1}{Previous File Hash (21 bytes)}
  \end{bytefield}
  \caption{Key packet byte structure: AES-256 encryption key, SN - Packet sequence number, RI - Client reconnect interval, Video ID for file retrieval, first 21 bytes of last recording SHA-256 file hash.}
  \label{fig:packet}
\end{figure}

The video file is encrypted with AES-256, a cryptographically secure encryption function and choice of key length. This video is uploaded to be available at the URL broadcast to listening devices and deleted locally. All video processing occurs in RAM to reduce the probability of file restoration or key recovery. The Bluetooth packet structure is outlined in Fig. \ref{fig:packet}.

The Camera Info Service provides meta-information, version information, a \textit{`friendly'} name and location details. The file location structure is also accessible here with the following structure:

\begin{center}
  \verb"<scheme>://<address-of-file-store>/<hex-encoding-of-video-id>.{mp4|jpg}"
\end{center}

\begin{figure}
  \includegraphics[height=180pt]{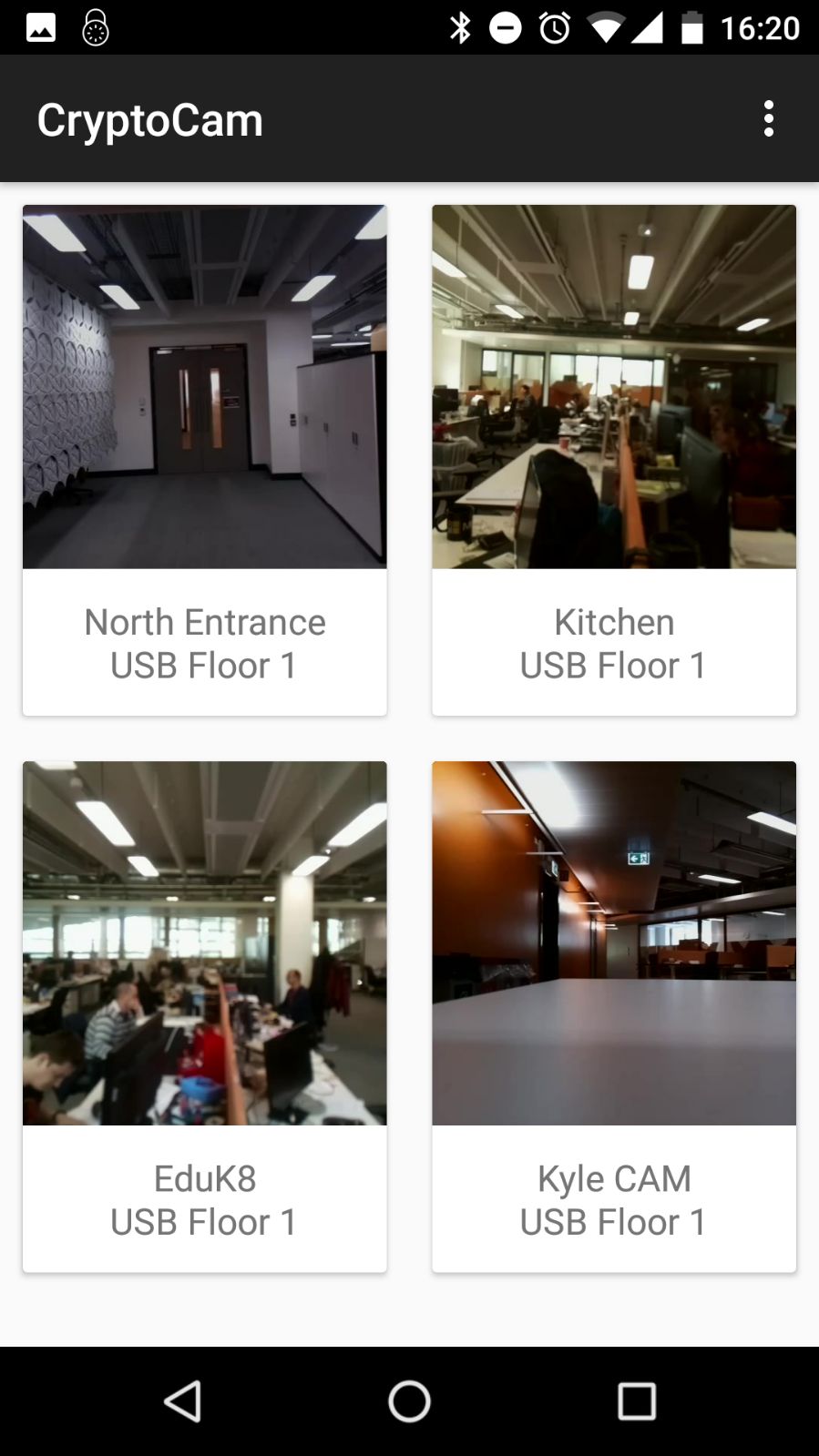}
  \includegraphics[height=180pt]{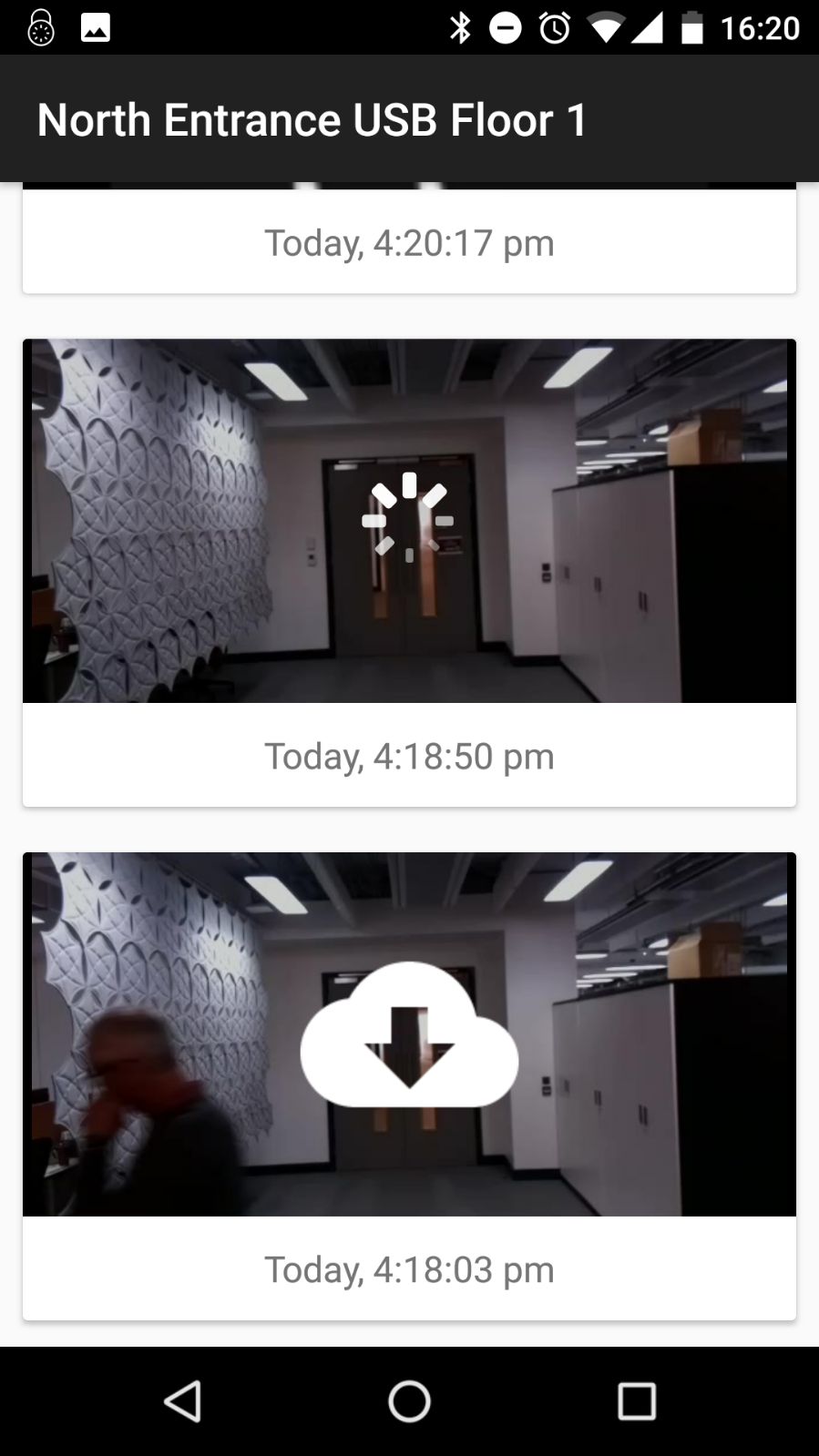}
  \includegraphics[height=180pt]{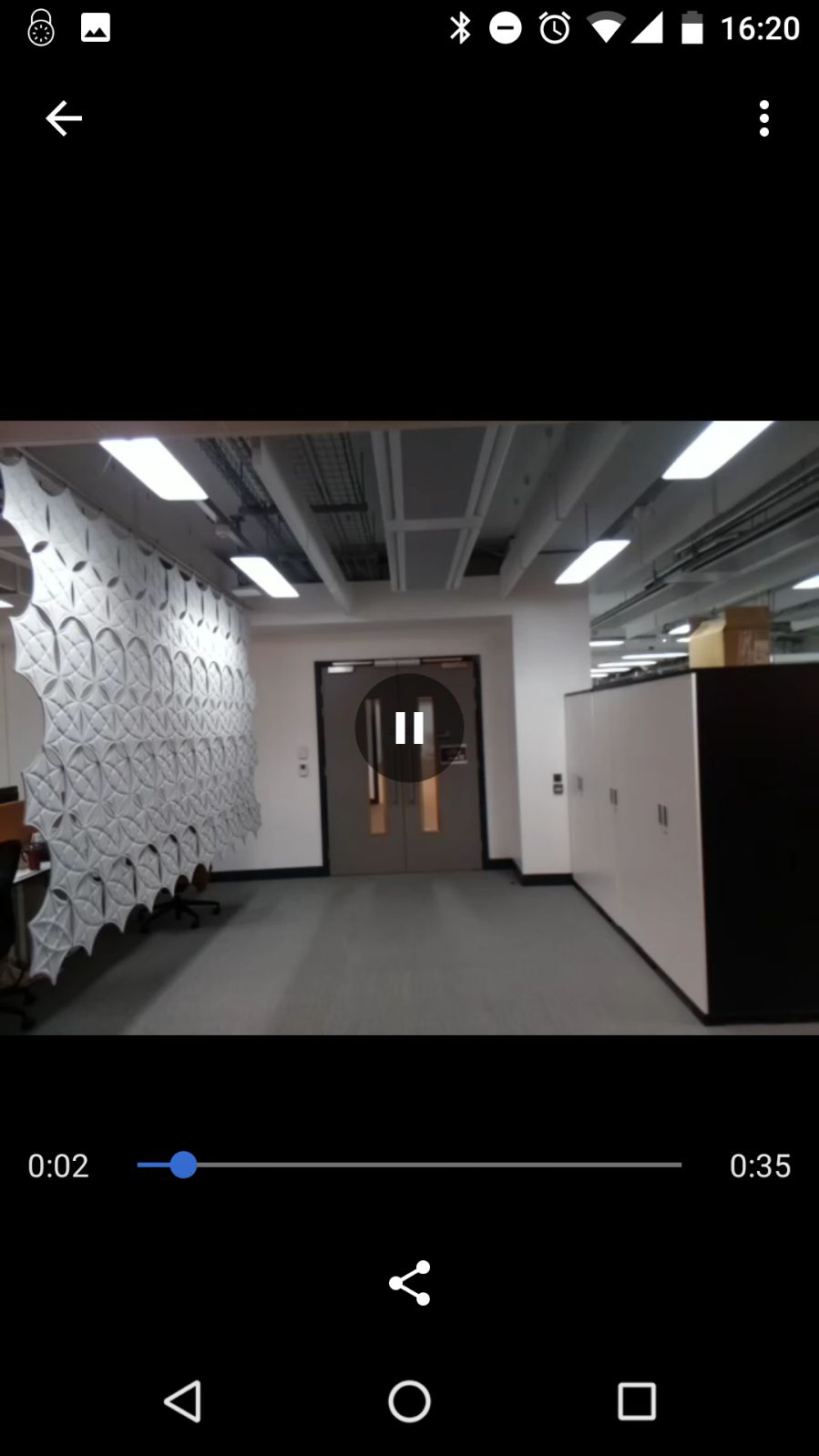}
  \caption{CryptoCam Android application.}
  \label{fig:ccapp}
\end{figure}

\subsection{Listening Client Implementations}
Listening clients run on a user's smartphone as a passive application in the background, collecting keys from nearby cameras. These keys are stored for later use within the application. Whenever the user wants to see when they have been the subject of a recording, the discovered cameras and video segments are organized into groups based on their proximity to one another. If the user requests to play a video from within the application, the appropriate encrypted file is retrieved from the URL originally provided by the camera. This file is then decrypted locally using the AES-256 key included in the original key packet, then played back. Implementations of the client have been made available for iOS\footnote{Apple iOS, https://www.apple.com/uk/ios/ios-11/} and Android\footnote{Google Android, https://www.android.com} though, due to background processing limitations on iOS, the key listener is currently less reliable on this platform.

\subsection{Finding Nearby Cameras}
Using local broadcast technologies we pose that CryptoCam can provide a reliable communication, not only of its presence in a area, but also further camera meta-data. CryptoCam advertises at regular intervals with Bluetooth Low Energy broadcasts, and can be interrogated by clients for further details. To address the discoverability of cameras users need to be able to reliably and easily discover when they were/are being recorded. Each potential subject of a camera in the immediate vicinity of a camera should be notified of its presence and perhaps, depending on the application scenario, its purpose. Other meta-information about the camera, how the data is handled, length of storage, audio recording, etc. are all important considerations for users near a particular camera. Using Bluetooth characteristics for static configuration variables, the camera configuration can be provided to users (outlined in Table \ref{tab:ccdescriptors}).

\begin{table}
\caption{CryptoCam Descriptors}
\label{tab:ccdescriptors}
\begin{minipage}{\columnwidth}
\begin{center}
\begin{tabular}{ccc}
  Characteristic 	& Options 							& UUID\\
  \toprule
  Name 				&  									& 0001\\
  Mode 				& \textit{auto, manual, delayed} 	& 0002\\
  Location 			& \textit{coordinates, description} & 0003\\
  URL Format 		& 									& 0004\\
  Key 				& 									& 0011\\
  \bottomrule
\end{tabular}
\end{center}
\bigskip\centering
\footnotesize
 All UUIDs are prefixed with the CryptoCam ID: cc92cc92-ca19-0000-0000-00000000\#\#\#\#.
\end{minipage}
\end{table}

\subsection{Simple File Access}
A right of access to footage is a crucial principle of CryptoCam. Simple, reliable exercising of this right with CryptoCam is achieved through being in the vicinity (Bluetooth range) of a camera and retrieving encryption keys. This provides a user's right to access, with the scope of their access pre-determined by the camera configuration. Using the encryption keys obtained from the camera at the point of capture, users can pull the relevant recording and decrypt its contents for local playback. This process provides a balance between secure and verifiable access, and simple and readily available access. Crucially, this process is automated, requiring minimal direct administration.

\subsection{Non-Exclusive Design}
The features of CryptoCam outlined in this paper aim to address some of social and procedural issues around CCTV and private camera operation. However, many of these features could arguably be considered to provide unacceptable restrictions for a camera operator. The features of CryptoCam have been designed to be non-exclusive, in that the level of implementation of each feature can be interrogated by client devices and differ based on operator preference.

\section{Study}\label{study}
To evaluate the concept of CryptoCam we constructed a user study to asses this challenge to existing CCTV infrastructure and perceptions. We chose a study that would not only highlight our approach to information sharing but also build a meaningful application of this footage for users. An obvious challenge to running a study with a technology such as this is that many of the scenarios and use cases we have discussed so far rely on ubiquity of CryptoCam cameras. As this was an insurmountable problem for our user study we decided to choose a constrained environment, the workplace. We deployed xx cameras around Newcastle University's Urban Science Building covering a range of spaces. Meeting rooms, desk spaces, kitchens and other general break out spaces were all covered. The CryptoCams deployed were all no audio cameras following the automatic upload and share procedure outline above. Cameras were set up as wide angles to avoid capturing detail and provide more of a contextual shot, reducing potential privacy issues with key bleed.

In order to provide a more compelling experience to our participants and demonstrate the possibilities for more open camera sharing policies we were also to construct films of participants day. As part of the study we requested access to meta-data about our participants. Namely their heart rate data (a fitness tracker was provided if they did not already have one), general fitness data and calendar. As we were deployed in a workspace we hoped to combine these data sets to produce compelling films of their day, integrating meta-data into the films to enable users to reflect. Essentially we wished to use the meta-data to index the footage, providing insight into this potential information overload. The study was run over the course of a week (5 working days).

We recruited 10 participants (avg.) to install the CryptoCam application on their Android phones (iOS system limitations restrict CC usage). The application provides basic viewing capabilities, collecting keys from cameras the users pass throughout the day. The participants were encouraged interact with the application regularly and reflect on the privacy implications of such a system contrasted with the freedom of access. We used a slightly modified client application to log the keys for footage to enable the next step of the study. At the end of the week we retrieved the logged keys from each of our participants phones along with their activity and calendar data for that week. We constructed short films of their day referencing significant portions of their day denoted by their activity or calendar data or a combination of both. Used to index against the CryptoCam footage keys we constructed a custom film of their week highlighting segments of CryptoCam film where their heart rate was elevated, they were exercising or they had a meeting near a camera.

Each participant was then presented with their film in an interview session. Each film was an average of xx mins long. After the participants had watched their film we conducted interviews with each participant. In the interviews we sought to discover what our participants perceptions of existing CCTV technology was, including their perceptions or perhaps experience of access procedures related to these cameras. We then sought reflections on CryptoCam's more open data sharing policies, whether technologies such as this could change their perceptions around CCTV, if this would make them more comfortable with cameras being a more ubiquitous environmental sensor in workplaces, homes and other public places. Finally we asked participants to reflect on the films we had created for them, if such an application of ubiquitous cameras also influenced their perceptions around CCTV, moving it beyond simply a security/protective device towards a more personal reflective device perhaps.

Participant interviews were semi-structured, with thematic ana

\subsection{Results}

\section{Discussion}\label{discussion}
CryptoCam provides the opportunity to democratize access to footage. Simple, passive and secure sharing of footage from any camera while potentially addressing pre-existing privacy issues relating to camera access also opens up new issues of privacy.  The concept of trust with CryptoCam is one of the system's most important concerns. The details of the key sharing, footage data storage and access, and the provenance of the data, are all intrinsically linked to the privacy features of the camera.  

\subsection{Bounded Key Transfer}
\begin{figure}
  \includegraphics[height=150pt]{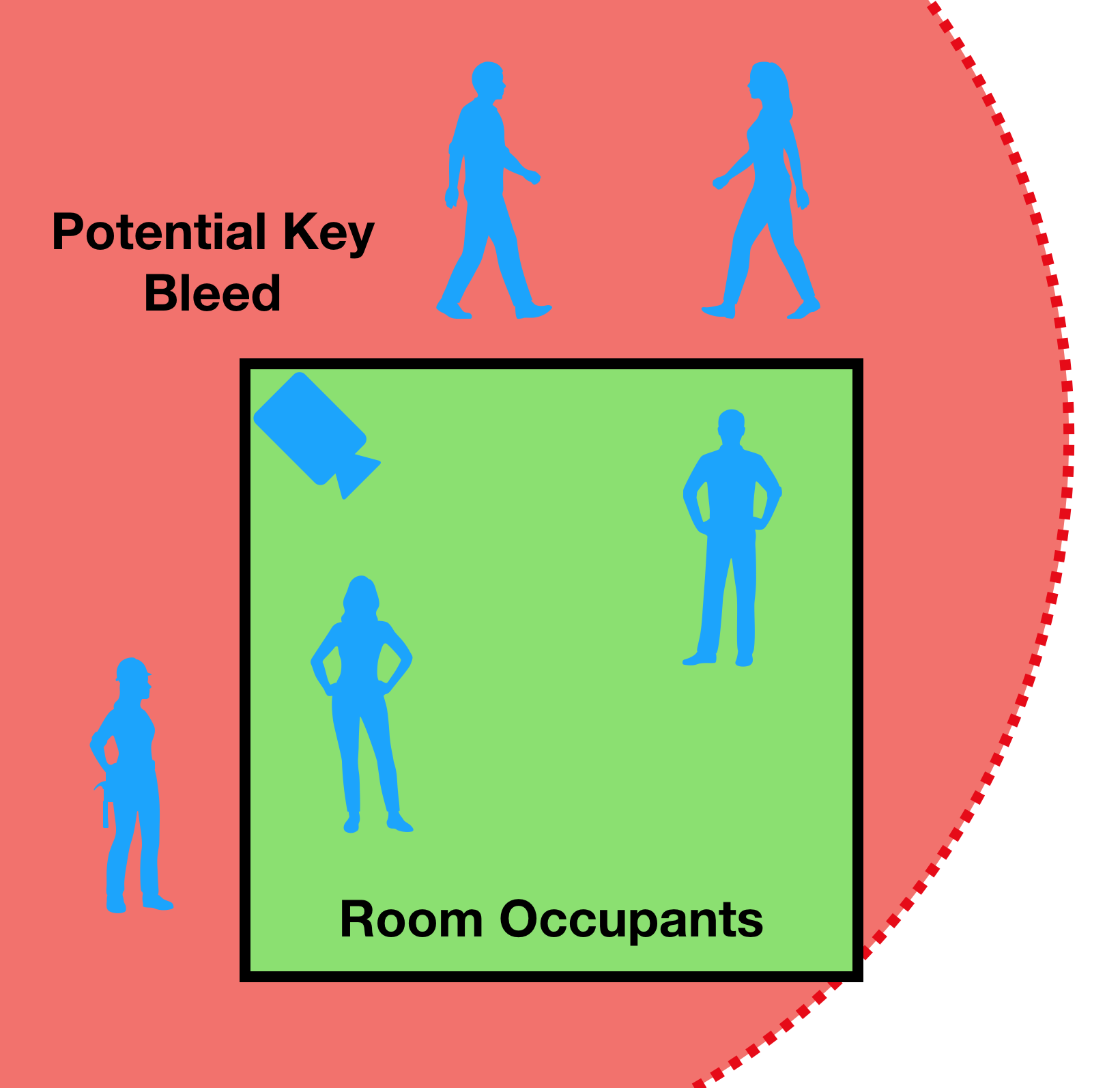}
  \caption{Diagram showing Bluetooth range exceeding the confines of the intended recipients, the occupants of the room being observed by the camera.}
  \label{fig:keybleed}
\end{figure}

Wireless transmissions cannot guarantee that coverage is exclusively bounded to a certain area. Key \textit{`leakage'} occurs where footage access tokens are provided to those who are not subjects of the recordings and are outside the locus of the surveillance area. See the example shown in Fig. \ref{fig:keybleed}, some recipients are not in the subject room. Techniques that may address this problem are outlined in this section.

\subsubsection{Explicit Verification Processes}
Proof of presence could be made more explicit. For example, by entering a randomly generated code shown on the camera at regular intervals, or presenting to the camera a QR code on the screen of a subject's smartphone, would provide the verification required to ensure that only subjects within the view of the camera can access footage. However, this technique is far more intrusive and inconvenient to users.

\subsubsection{Vision Analysis}
The possible key leakage issues we envisage could may be solved by matching an individual in the footage using Computer Vision techniques.  A reliable means of identification of which key listening device corresponds to which individual subject within the frame, would ensure that only those who should get access to footage are given access. Several techniques have been explored in existing literature for identification of people from video, including: association of accelerometer data from phones with detected walking gate \cite{Nikou2016CCTVWalk}, identifying users with a view to obscure them from images \cite{Zhang2016Privacy-friendlySystem} and other facial tracking work \cite{McKenna:2016:APS:2890602.2890612}. These techniques demonstrate the possibilities for identification of users however, they require some corresponding information from subjects. This information is in some cases personal or requires sacrificing other non-identifiable information \cite{Zhang2016Privacy-friendlySystem}. CryptoCam as a concept was developed to preserve the privacy of the subject, it does not require any user login or other identifying process, introduction of any such features should be carefully balanced with any related privacy sacrifices.

\subsubsection{Audio Chirps}
Audio \textit{`chirps'} are a well used method of providing a unique and discrete means of passing digital data between devices through audio channels. The benefit of this method of communication for applications such as CryptoCam is that audio chirps are more localized and that a chirp is less likely to go beyond the bounds of a room, for instance. However, access to the background audio processing required to achieve this is often blocked on mobile devices. Furthermore, the range of audio chirps sent or received from a smartphone may be broader or narrower than the intended range, as such this technique of verification should be used in conjunction with others. Amongst other techniques, Google's Nearby API\footnote{Google Nearby API, https://developers.google.com/nearby/} deploys audio chirps, and demonstrates the technical feasibility of this approach to proximity verification.

\subsection{Access Granularity}
\subsubsection{Levels of Access}
Footage from CryptoCam could have levels of access described to preserve the privacy of potential subjects. Altering the resolution, regions of footage or audio according to some calculated level of trust for a key listening client could address some of the concerns around privacy in access. Used in conjunction with other techniques outlined above, a user could have a base access level with a low resolution version of the footage. When a particular user is validated as being within the scene (for instance, from vision analysis) they could be provided with keys to the high resolution version. When a phone is close enough to be identified (for instance, with an audio chirp from the phone or camera) then the audio track could be provided to the user. This technique of discrete levels of access to footage could limit the potential impact of errors in determining subjects, whilst still providing simple and unobtrusive access.

\subsubsection{Advertise Tokens for Fine-Grained Recording Intervals}
To reduce wireless network contention, and better manage limited resources, CryptoCam allows for pre-configuration of the recording interval to provide appropriate time intervals for client connection based upon expected size of audience. A large set of subjects needing to connect and read keys from a camera requires a longer interval to ensure keys are not missed (due to saturation of the channel). However, a longer interval leads to more information bleed: i.e. if a subject is only present for a fraction of an interval they receive footage for the entire interval. Using Bluetooth advertising packets, which are broadcast, we can potentially address this issue. By advertising tokens at a much shorter interval, although these tokens alone cannot be used to decrypt footage, a final key can be retrieved from the camera which decrypts the file into a selection on smaller chunks to be decrypted with the available tokens. Such an implementation provides potential for much smaller segment lengths with a lower risk of information bleed.

\subsection{Distributed File Access}
Footage ownership remains a significant issue in any democratized video access protocol. Sharing content from cameras relies on the camera operator making this footage available to others. We propose that subjects have a right of ownership to recorded footage from cameras, de-centralizing the storage and ownership model of cameras potentially has benefits for both camera operators and subjects, breaking down the current asymmetric ownership model of CCTV and private cameras. Deploying technologies such as IPFS, built upon the principles of peer-to-peer file sharing, content can be distributed to many devices for later reconstruction. Operators can potentially save storage space by distributing their recording amongst subjects with a means to access these files at a later date if necessary. Subjects can have direct ownership over the recordings with these files stored on their devices or have access, equal to that of the operators, to a shared recording.

\subsection{Hardware/Software Verification}
To ensure footage is handled as outlined in this paper we propose a verification process for both software and hardware solutions. Laptops and smartphones are typically privacy protected through hardware (e.g. camera light illuminated when active) and software (e.g. explicit access prompts) means. Similarly, we need a means to verify that the CryptoCam camera feed is being handled in the way it is intended. Footage from the camera should be encrypted at point of capture and stored in memory to reduce the likelihood of physical access recovery of unencrypted footage. Trust in third-party developed cameras could be improved by using hardware camera modules which encapsulate this process, including the broadcast of keys. Authentication chips similar to those present in Apple's MFi\footnote{Apple MFi Program, https://developer.apple.com/programs/mfi/} program products and Trusted Platform Modules\footnote{Trusted Platform Module, https://docs.microsoft.com/en-us/windows/security/hardware-protection/tpm/trusted-platform-module-overview} could ensure that the camera has not been tampered with. Software solutions may also be required (e.g. a smartphone application), these solutions could also be verified to ensure compliance with the CryptoCam standard. Such practices are widespread in various industry technologies, CryptoCam particularly mandates such a process to ensure trust in the system. The level of CryptoCam implementation can also be encoded into this verified camera process, CryptoCam's non-exclusive implementation allows for features to be implemented to varying degrees of completeness. Providing a clear representation of which components are private and which have been altered is crucial to build trust in the system.

\subsection{Data Protection Regulation}
Data protections issues, in particular the European General Data Protection Regulation, is a topic of great concern to industry at the moment. GDPR and CryptoCam come from similar motivations and backgrounds. Moving the locus of control over data from corporations towards the subjects whom the data is about. We argue that many of the subsequent realities of a deployment of CryptoCam satisfy legislation and CryptoCam has the potential to be a powerful tool for both subjects and operators, with compliance with GDPR one of many benefits alongside the potential for distributed storage and streamlined handling of access requests to footage. With data access being a key part of GDPR, a means to expedite these processes would be a boon to camera operators.

\subsection{A New Dawn for CCTV}
CryptoCam is an incredibly flexible concept, building out from the Open Circuit Television framework introduced above, we address many of the issues highlighted. The flexibility of the concept allows for deployment in appropriate situations with appropriate configuration. We posit that technology solutions such as these provide a new means for configuration for privacy and access for organizations. Applying the principles of the OCTV framework in conjunction with novel technology concepts such as CryptoCam we hope to provide a system of security and surveillance cameras which is more open and subject to scrutiny. Redressing the balance of power over footage with concepts such as CryptoCam we hope to change perceptions about surveillance cameras in our lives, restoring them as protective devices.

\subsection{CryptoX}
CryptoCam has been discussed as a video camera, however the concept is not intrinsically linked to video alone. As such we also propose an expansion of this concept, something we call CryptoX. Sensor data, presence proof, audio and photos can also be used with CryptoX. For example, sensor data and audio collection of this information by a local government or private organization. As a subject of this data you, perhaps, should reasonably have access to this data. Also, discovering what is being collected and why, with a system to discover the sensor equipment deployed nearby. CryptoX can also be used for proof of presence, large parts of our legal processes are still archaic in that they rely on a completely un-verifiable signature. Making digital signing a more convenient and provable process with proof of presence mechanisms supported by CryptoX could revolutionize many aspects of legal processes.

\section{Conclusion}\label{conclusion}
The concept Open Circuit Television (OCTV) has potential to be an important step in respect of ubiquitous camera systems, offering a novel route towards configuring a camera system that is more concomitant with the needs of those who are likely to be recorded. We believe this will enhance public trust in public camera settings, enable or smooth their deployment in sensitive settings (especially in the context of data protection legislation), and democratize the process of accessing camera footage, shifting the emphasis from the watcher to the watched. Through CryptoCam itself, we have provided a pragmatic implementation of OCTV, which we are making available to the community to adapt for camera systems.\footnote{We will publish the Open Source version of our implementation with this paper (and which will be linked here): this is currently omitted for the purpose of blind review.}  

The initial motivation and overall developed goal of this project was to increase trust in CCTV. We deploy the term Open Circuit Television as a provocative challenge to open access to these typically closed (or rather, obscure) systems. We hope to rebuild trust in cameras as a protective device through proposing a system that prohibits typical abuses of existing cameras whilst also presenting novel and interesting new discovery and playback experiences for users, and as a potentially powerful tool in our everyday lives. Google Clips\footnote{Google Clips, https://store.google.com/?srp=/product/google\_clips} and other life-logging cameras\footnote{Narrative Clip 2, http://getnarrative.com/} demonstrate an industry and consumer interest in collecting more media in our lives, however, this needs to be weighed against privacy issues that may arise from this additional capture.

We consider that this framework will be of wide utility: there are all manner of scenarios beyond fixed cameras where CryptoCam (or an implementation thereof) would be beneficial. Whilst our framework, as articulated in this paper, is focused upon fixed cameras (for ease of description), we note that CryptoCam could be easily used in a mobile setting, where wearable or smartphone mounted camera footage is made available to people who were within their vicinity. This would range from the social (e.g. gathering footage in a manner akin to Bootlegger \cite{Schofield2015Bootlegger} at a public event or concert: the difference would be that anyone could control the footage), onto police body-cams and individuals engaged in the practice of sousveillance \cite{Mann2002Sousveillance:Environments.}. Similarly, the broader framework need not be constrained to the transmission of video: other data could also be transmitted and distributed using the CryptoCam mechanism, including sensor data, photos, audio, documents, and for any scenario that would benefit from location-orientated access control. Although this is a matter for future work, all of these settings represent a fruitful opportunity to explore the usage of CryptoCam, either as a discrete system or a component of another camera implementation, fitting with the existing agenda surrounding cameras and Ubicomp.

We hope that a renewed focus on openness in public cameras will lead to a more trusted and socially acceptable set of security cameras. Bringing cameras back to their original purpose as a deterrent and protective device for the public, rather than a measure for \textit{`Big Brother'}.

\bibliographystyle{ACM-Reference-Format}
\bibliography{Mendeley}

\end{document}